\documentclass{elsart}
\usepackage{amssymb}
\usepackage{graphicx}
\usepackage{color}

\begin{document}
\begin{frontmatter}

\title{Storage of very cold neutrons in a trap with nano-structured walls}

\author[JINR]{E.~V.~Lychagin}
\author[JINR]{A.~Yu.~Muzychka}
\author[ILL]{V.~V.~Nesvizhevsky}
 	\ead{nesvizhevsky@ill.fr}
\author[LPSC]{G.~Pignol}
\author[LPSC]{K.V.~Protasov}
\author[JINR]{A.~V.~Strelkov}

\address[ILL]{ILL, 6 rue Jules Horowitz, Grenoble, France, F-38042}
\address[LPSC]{LPSC (UJF, CRNS/IN2P3, INPG), 53, rue des Martyrs, Grenoble, France, F-38026}
\address[JINR]{JINR, 6 Joliot-Curie, Dubna, Moscow reg., Russia, 141980}

\date{\today}

\begin{abstract}
We report on storage of Very Cold Neutrons (VCN) in a trap with walls containing powder of diamond nanoparticles. 
The efficient VCN reflection is provided by multiple diffusive elastic scattering of VCN 
at single nanoparticles in powder. 
The VCN storage times are sufficiently long for accumulating large density of neutrons 
with complete VCN energy range of up to a few times $10^{-4}$~eV. 
Methods for further improvements of VCN storage times are discussed.
\end{abstract}


\end{frontmatter}

\section{Introduction}

Recently we showed that powders of nanoparticles could be used efficiently as first reflectors of Very Cold Neutrons (VCN) 
in the complete VCN energy range \cite{Nesvizhevsky2008}, 
thus bridging the energy gap between efficient reactor reflectors \cite{Fermi} for thermal and cold neutrons, 
and effective Fermi potential for ultracold neutrons (UCN) \cite{Shapiro}. 


The use of nanoparticles provides a sufficiently large cross-section for coherent scattering 
and inhomogeneity of the moderator/reflector density on a spatial scale of about the neutron wavelength 
\cite{surfacenanoparticles2}. 
A large number of diffusive collisions needed to reflect VCN from powder constrains the choice of materials: 
only low-absorbing ones with high effective Fermi potential are appropriate. 
Thus, diamond nanoparticles were an evident candidate for such VCN reflector. 
The formation of diamond nanoparticles by explosive shock was first observed more than forty years ago \cite{Carli}. 
Since then very intensive studies of their production and of their various applications have been performed worldwide. 
These particles measure a few nanometers; 
they consist of a diamond nucleus within an onion-like shell with a complex chemical composition \cite{Aleksenskii}. 
A recent review of the synthesis, structure, properties and applications of 
diamond nanoparticles can be found in \cite{Dolmatov}.

The first experiments on the reflection of VCN from nano-structured materials as well as on VCN storage were carried out in the seventies in \cite{Steyerl} and later continued in \cite{Arzumanov}. 
In \cite{Nesvizhevsky2008} we extended significantly the energy range and the efficiency 
of VCN reflection by exploiting diamond nanoparticles. 
A reflector of this type is particularly useful for both UCN sources using ultracold nanoparticles 
\cite{surfacenanoparticles2,moderator} and for VCN sources; 
it would not be efficient however for cold and thermal neutrons, as shown in \cite{Artemiev}.

In order to measure precisely the VCN reflection probability from powder of diamond nanoparticles and to explore feasibility of VCN storage in traps with nano-structured walls, 
we carry out a dedicated experiment described in the present article. 
In fact, the measuring procedure used here is equivalent to that typically used in experiments on UCN storage in traps (see for example \cite{Ignatovich,Golub}). 
The difference consists in a type of trap walls and in characteristic values of the reflection probability. 
In the section 2 we describe the experimental setup. 
In the Section 3 we present the experimental results, and analyze them in the Section 4.

\section{The experimental setup}

\begin{figure}
\begin{center}
\includegraphics[width=1.\linewidth]{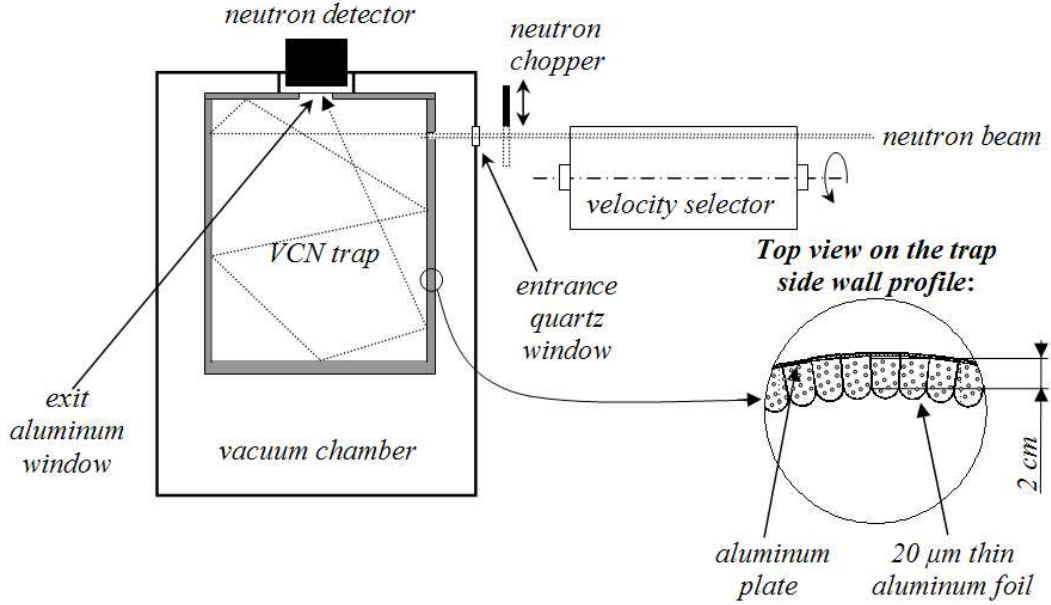}
\end{center}
\caption{The installation scheme.} \label{scheme}
\end{figure}

This experiment was carried out at the VCN beam, PF2, ILL. 
The installation scheme is shown in Fig. \ref{scheme}.

The VCN trap has cylindrical shape with a diameter of 44~cm and a height of 47~cm. 
VCN could enter the trap through a small square window of 2~cm by 2~cm in its side wall. 
The VCN beam diameter is $\approx$1~cm. 
VCN could be reflected many times from the trap walls. 
Thus they could find an exit circle window with a diameter of 6~cm in the trap cover and enter into a detector behind the window. 
The VCN beam could be opened or closed using a fast cadmium valve with a thickness of 0.2~mm. 
The VCN velocity could be chosen using a velocity selector in front of the valve. 
The trap is placed inside a vacuum chamber with an entrance quartz window with a thickness of 3~mm and 
an exit aluminum window with a thickness of 1~mm. 
When the VCN beam is closed the detector count rate decreases exponentially following the VCN density in the trap. 
Thus we could measure the VCN storage times as a function of their velocity. 

The neutron detector is a gaseous proportional $^3$He counter 
(the thickness of its sensitive layer is 5~cm, the $^3$He partial pressure is 200~mbar) 
with an entrance aluminum window with a thickness of 100~$\mu$m and a diameter of 9~cm. 
The detector electrical signals are analyzed using the time and amplitude analyzes. 

The velocity selector consists of a cylinder with a length of $l=40$~cm and a diameter of $D=19$~cm. 
Curved plastic plates with a thickness of 1 mm are installed at the side surface of the cylinder in such a way that they form 
screw-like slits with a width of $d \approx 4.5$~mm and the screw length of $L=480$~cm. 
The cylinder rotates around its axis with the period $T$. 
Neutrons could scatter at hydrogen atoms in the plate's material; 
those neutrons leave the neutron beam. 
The rotation period defines the velocity of neutrons passing through the selector. 
Neutrons with low angular divergence and with momentum parallel to the selector axis pass through the selector screw slits, 
thus avoiding scattering by the plates only if the neutron velocity is equal to
\begin{equation}
v = L \frac{2 \pi}{T} \left( 1 \pm \frac{d L}{2 \pi D l}  \right) 
\end{equation}

\begin{figure}
\begin{center}
\includegraphics[width=.9\linewidth]{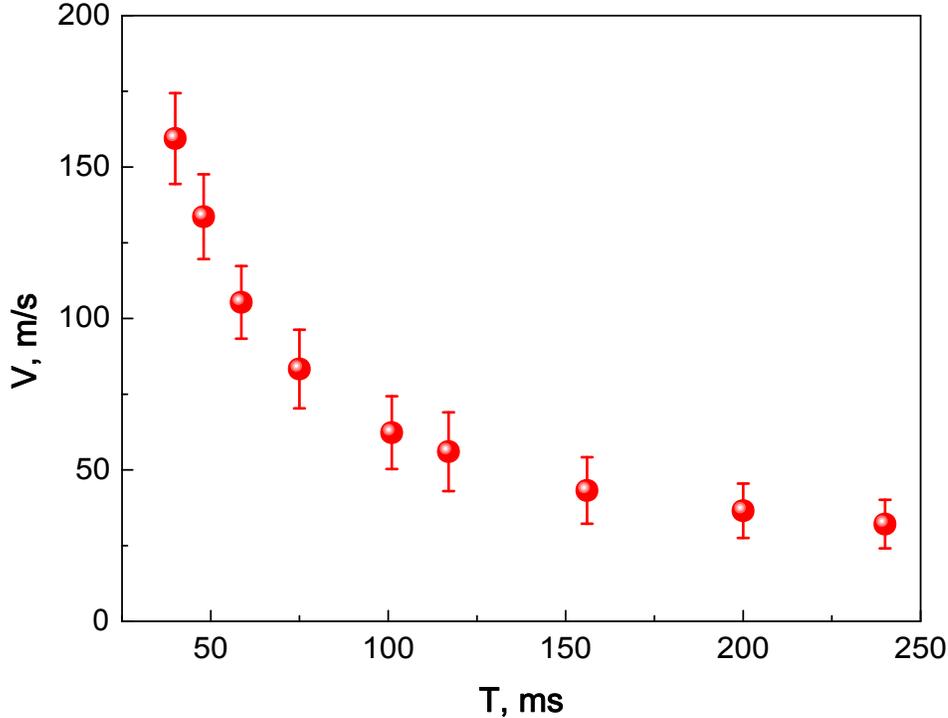}
\end{center}
\caption{The neutron velocity and at the velocity selector exit is shown 
as a function of its rotation period.
Error bars indicate the half width of velocity distribution in the beam after selector.
} \label{velocity}
\end{figure}

\begin{figure}
\begin{center}
\includegraphics[width=.9\linewidth]{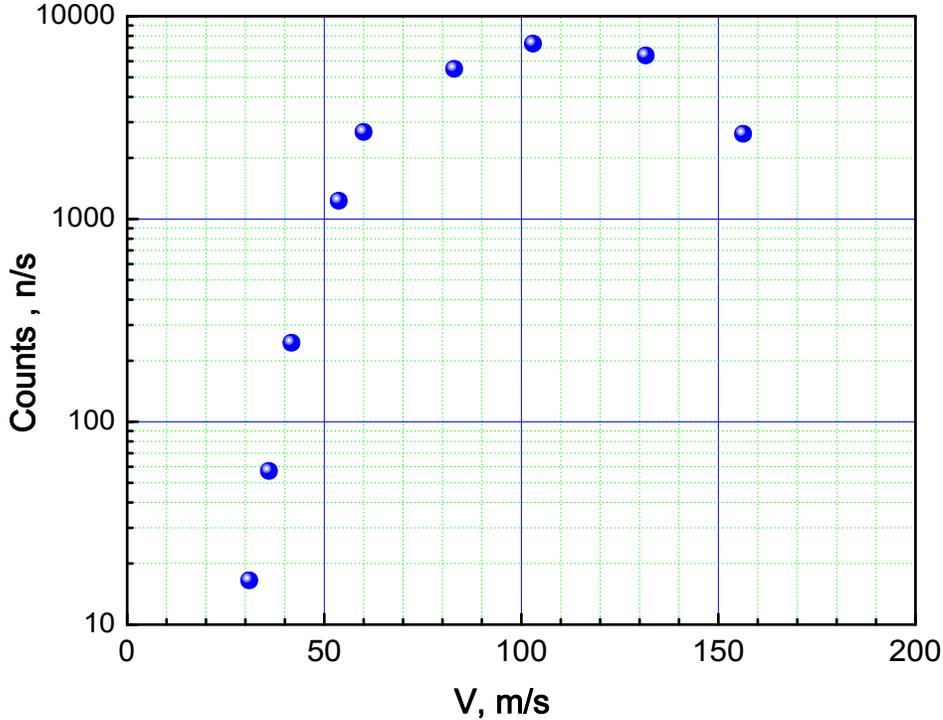}
\end{center}
\caption{The neutron flux at the selector exit is shown as a function of neutron velocity.} 
\label{fluxv}
\end{figure}

The selector allowed us to choose the neutrons with a velocity in the range of 30-160~m/s. 
The velocity resolution was measured using time-of-flight method; 
the result is shown in Fig. \ref{velocity}. 
It is equal to 10\% if $v=160$~m/s, and 25\% if $v=30$~m/s. 
The neutron flux at the selector exit is shown in Fig. \ref{fluxv}.

The valve is controlled by an electro-magnet governed with an electric pulse generator. 
The time of opening and closing the valve was measured in a separate experiment with a light beam; 
it is equal to 5~ms in both cases.

\begin{figure}
\begin{center}
\includegraphics[width=.9\linewidth]{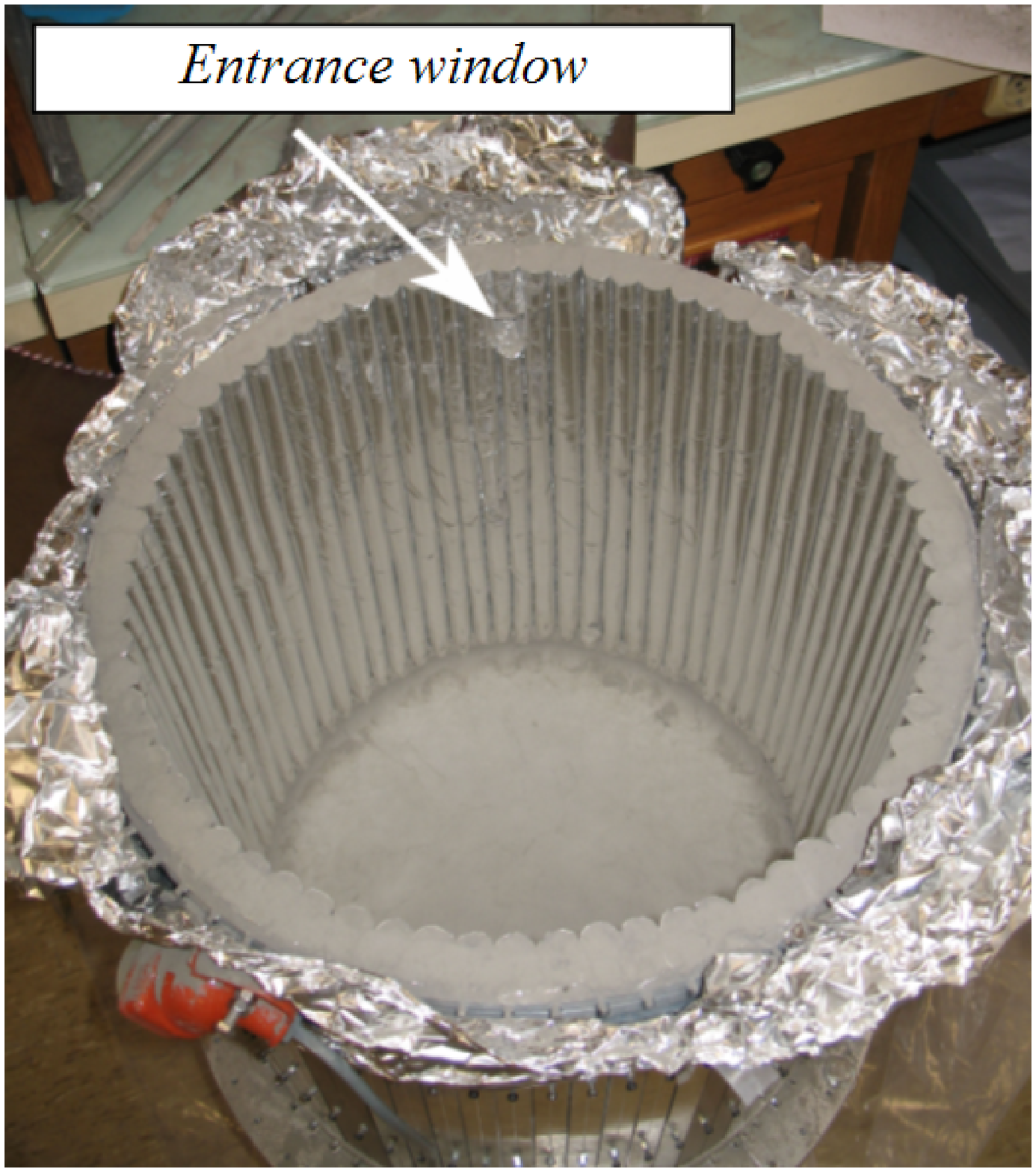}
\end{center}
\caption{The VCN trap. The cover is open.} \label{trap}
\end{figure}

\begin{figure}
\begin{center}
\includegraphics[width=.9\linewidth]{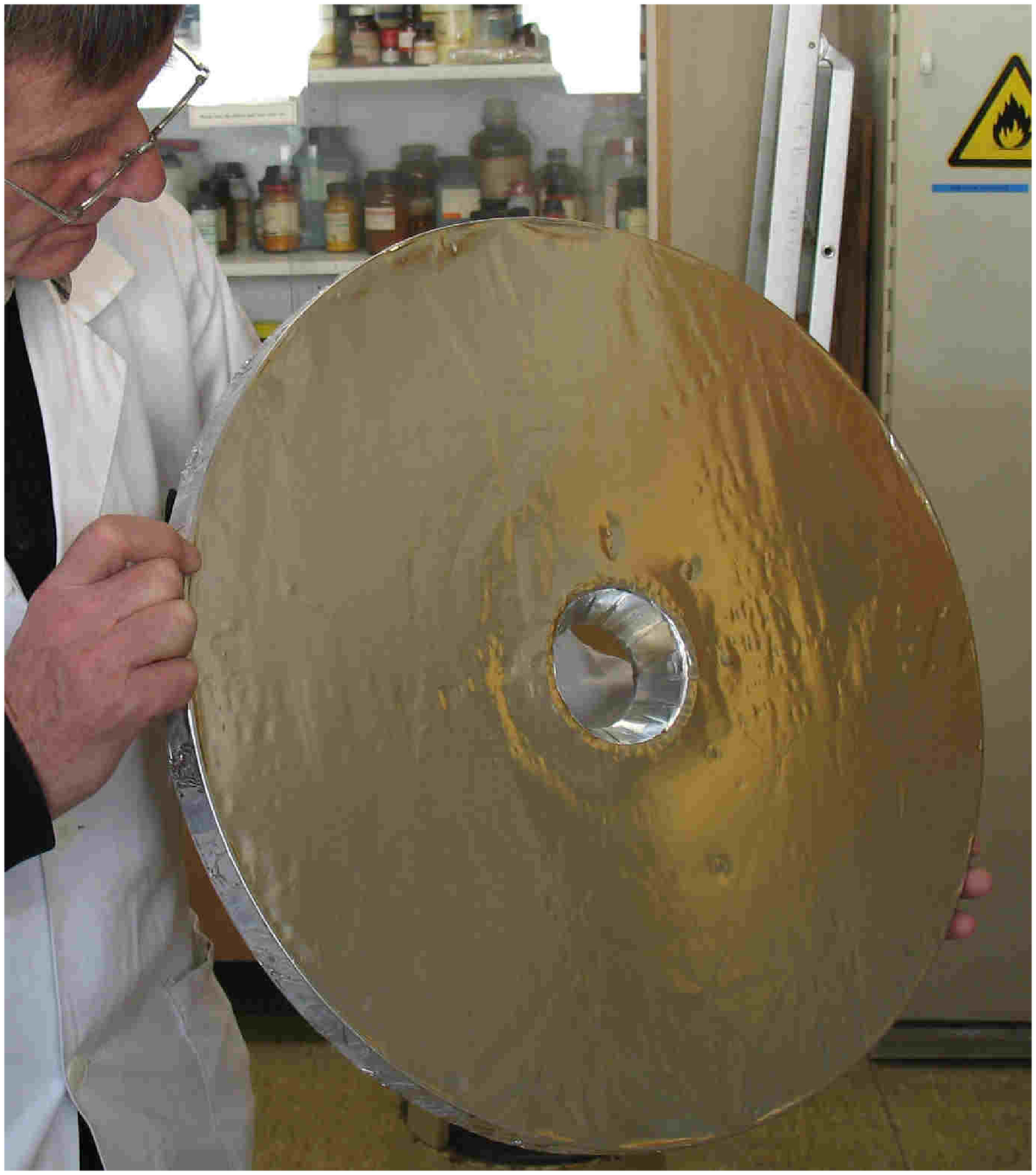}
\end{center}
\caption{The cover with a window in its center.} \label{cover}
\end{figure}

The trap is built using powder of diamond nanoparticles produced in accordance with standard TU-2-037-677-94. 
A complex technological problem consists in constructing vertical trap walls. 
We explored feasibility to make sufficiently rigid large "blocks" of nano-powder with open surface, which could be installed vertically. 
In particular, the nanoparticles were suspended in various liquids; 
then a liquid is evaporated. 
Thus, nanoparticles stick strongly to each other and form a kind of "blocks". 
However, a very significant change in density of nano-powder during this procedure damages such "blocks" and results to numerous cracks. 
Therefore we decided to fill nanoparticles into aluminum tubes, and to assemble them into a cylinder in order to get the side trap wall as shown in Fig. \ref{trap}. 
VCN reflects many times from powder; 
thus they pass many times through the aluminum walls. 
In order to avoid significant losses of VCN in aluminum, we have to decrease the aluminum wall thickness to a minimum value. 
Therefore the tubes are built using 20~$\mu$m thick aluminum foil rolled around aluminum plates 
as shown in Fig. \ref{scheme} and in Fig. \ref{trap}. 
The plate's width and thickness are 2~cm and 2~mm, respectively; 
the length is equal to the trap height. 
These plates form a cylinder, as shown in Fig. \ref{scheme} and in Fig. \ref{trap}. 
We fill in the tubes with powder, assemble the tubes, and compress powder in order to avoid eventual slits between the tubes. 
The compressed powder density is 0.4~g/cm$^3$. 
Stripes of filter paper are inserted between the plates and powder in each tube in order to facilitate pumping of air out of powder. 
The trap cover is shown in Fig. \ref{cover}. 
It consists of an aluminum disk, an aluminum circle with a height of 3~cm fixed to its perimeter, 
a foil with a thickness of 30~$\mu$m attached as shown in Fig. \ref{cover}, 
and a 3~cm thick powder layer over the aluminum foil. 
The powder density in the trap cover is 0.3~g/cm$^3$. 
A window in the cover center (see Fig. \ref{cover}) allows us to count VCN in the detector. 
The window side walls are produced of an aluminum foil with a thickness of 200~$\mu$m. 
The cover is installed on top of the trap side wall only when the upper edge of the compressed powder in the tubes is flattened as shown in Fig. \ref{trap}; 
thus an eventual slit between the side wall and the cover is minimized. 
The trap bottom is filled in with powder with a thickness of 3~cm compressed to a density of 0.3~g/cm$^3$.

Electro-heaters are rolled around the trap; 
the trap and the electro-heaters are covered by many aluminum foil layers in order to provide thermal isolation of the trap. 
The trap temperature could be raised using the electro-heaters and measured with two thermo-pairs attached to 
the massive aluminum parts of the trap bottom and the cover respectively.

\section{The experimental results}

When the valve is open VCN fill in the trap as long as needed to reach a saturation VCN density. 
Then we close the valve and measure the time constant of exponential decrease of the VCN density in the trap. 
The detector count rate is proportional to the VCN density in the trap. 
Such a measuring cycle is repeated many times in order to accumulate sufficient statistics. 

\begin{figure}
\begin{center}
\includegraphics[width=.9\linewidth]{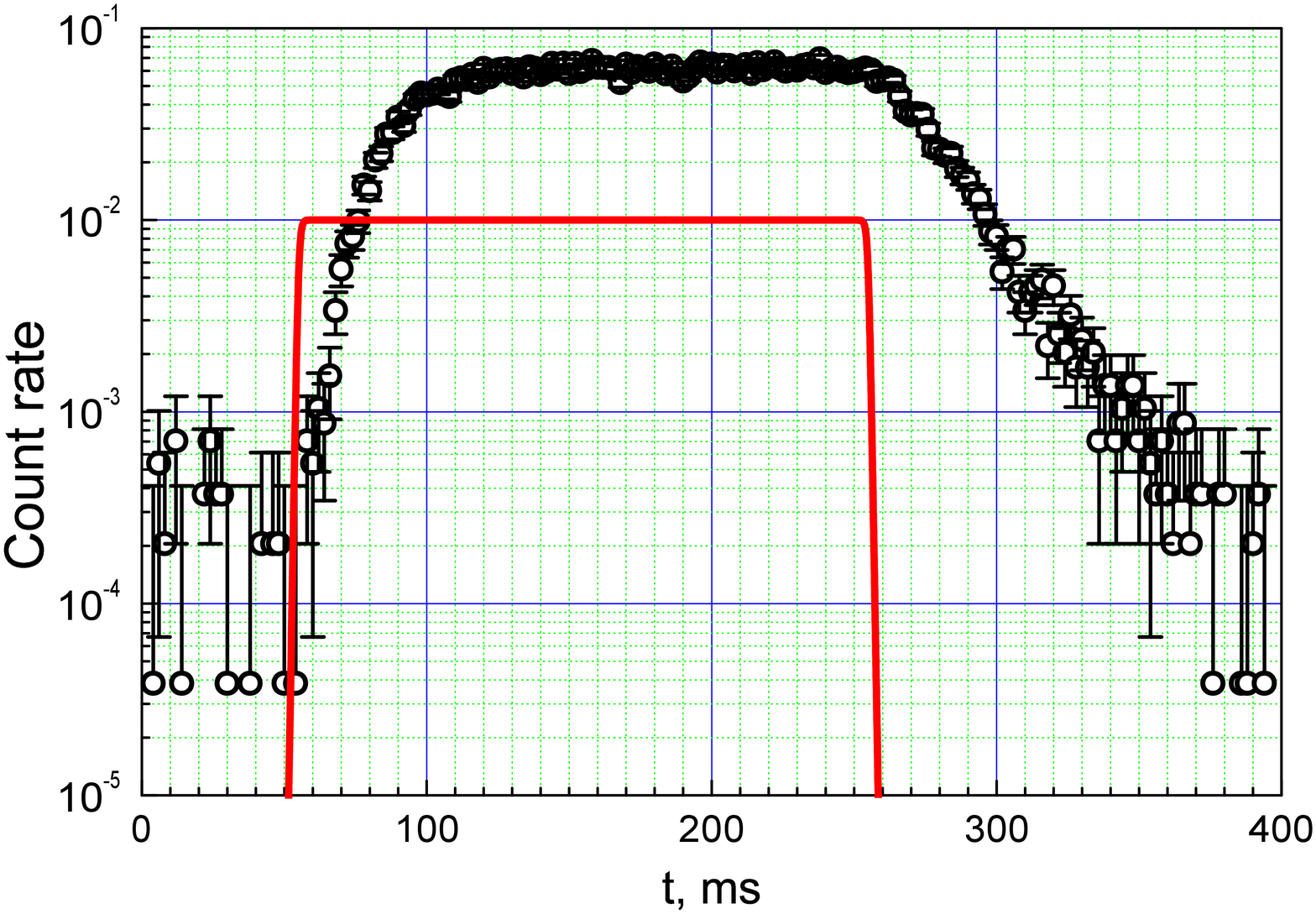}
\end{center}
\caption{An example of the neutron count rate during a measurement cycle for the VCN velocity of 85~m/s.
The solid line shows the light signal, in arbitrary units, used to synchronize the valve with the measurement cycle.} 
\label{counts}
\end{figure}

An example of the neutron count rate during one cycle is shown in Fig. \ref{counts}; 
a background is subtracted. 
The solid line in Fig. \ref{counts} shows the light signal, in arbitrary units, 
used to synchronize the valve with the measuring cycle. 
A pulse generator starts the cycle at time $t=0$, opens the valve at $t=50$~ms, and closes the valve at $t=250$~ms. 
When the valve is open the trap could be filled in with VCN up to a saturation density. 
When the valve is closed, the VCN density decreases exponentially. 
The characteristic time of this decrease is equal to the VCN storage time. 
The dominant loss factor is associated with the trap walls, as the area of the entrance and 
exit windows is smaller than 0.3\% compared to total area of the trap surface. 
The storage time in Fig. \ref{counts} is equal to $\tau^{VCN}_{st} = (19.0 \pm 0.5)$~ms.

Analogous measurements were carried out for this and other VCN velocities in 3 series of measurements. 
First, we pumped out the trap during 12 hours then measured the VCN storage times at room temperature. 
Second, the vacuum chamber was filled in with argon at a pressure of 1 bar then the trap was heated to a temperature of $\approx 120^{\circ}$~C and kept at this temperature during 3 hours. 
After that argon was pumped out, the trap was slowly cooled to a room temperature then VCN storage times were measured. 
Third, we heated the trap to a temperature of $\approx145^\circ$~C on permanent pumping. 
The measurement started when the temperature (measured with the two termo-pairs) had got stabilized. 
Besides, the detector count rate (with the valve open and the neutron velocity equal to 160~m/s) had to get stabilized as well. 
The later process is delayed because of slow heating of internal zones of powder.

\begin{figure}
\begin{center}
\includegraphics[width=.9\linewidth]{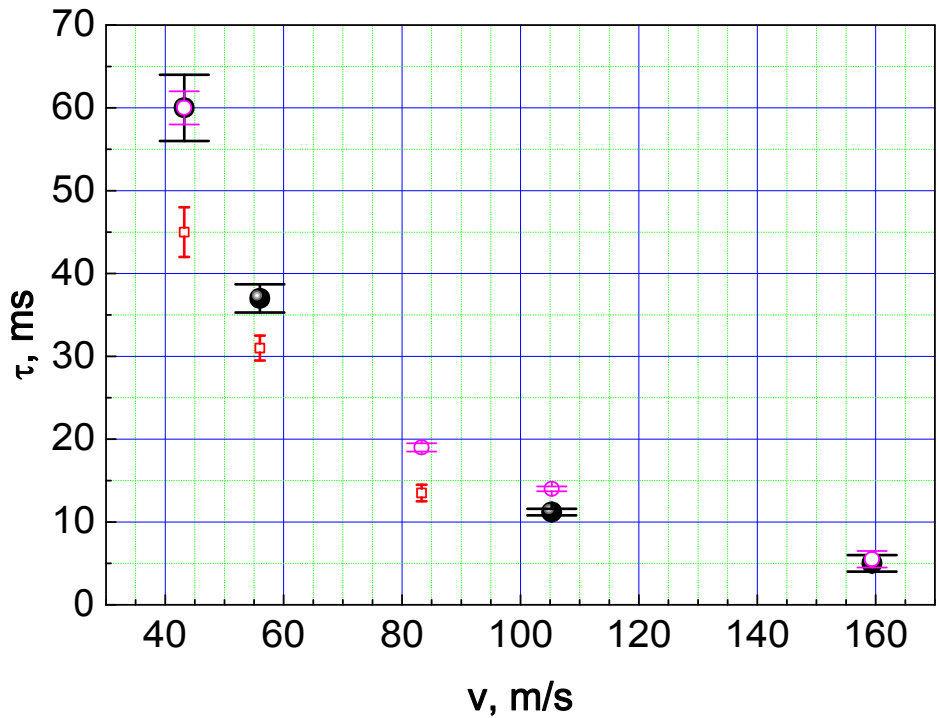}
\end{center}
\caption{The VCN storages times as a function of their velocity.
Black circles correspond to measurements at room temperature after 12 hour pumping. 
Empty circles show measurements at room temperature after heating the trap at $120^\circ$~C in argon. 
Boxes indicate results obtained at a temperature of $150^\circ$~C under permanent pumping.
} \label{storageTime}
\end{figure}

The results of all these 3 series of measurements are shown in Fig. \ref{storageTime}.

\section{Analysis of the experimental results}

\begin{figure}
\begin{center}
\includegraphics[width=.9\linewidth]{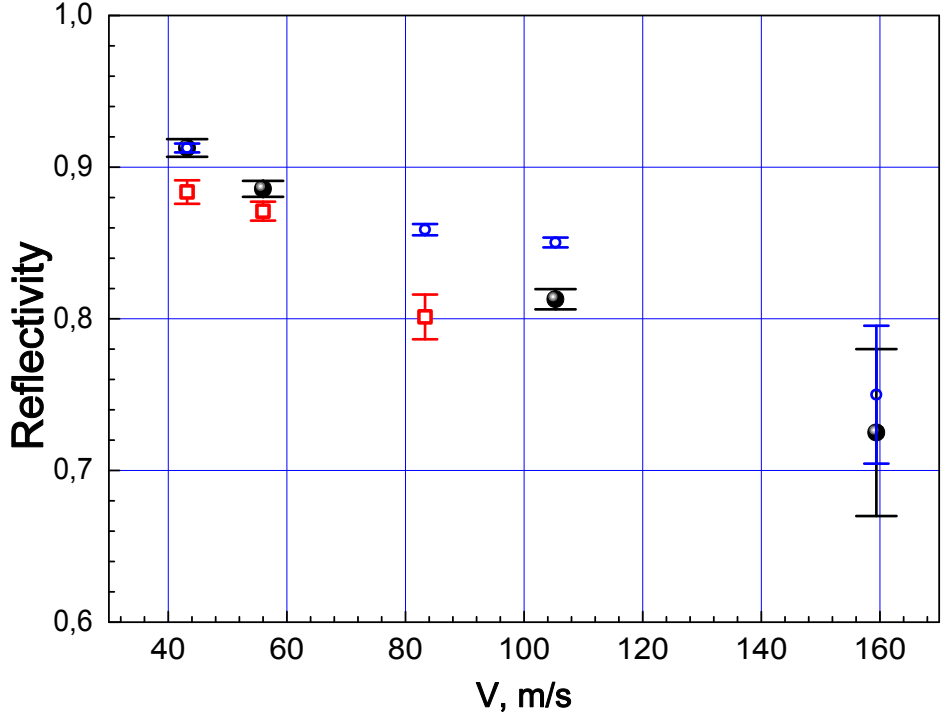}
\end{center}
\caption{
The probability of VCN reflection from the trap wall as a function of their velocity.
Black circles correspond to measurements at room temperature after 12 hour pumping. 
Empty circles show measurements at room temperature after heating the trap at $120^\circ$~C in argon. 
Boxes indicate results obtained at a temperature of $150^\circ$~C under permanent pumping.
} \label{reflectivity}
\end{figure}

For the trap geometry described above we get VCN mean free path for all neutron velocities equal :
\begin{equation}
\Delta x = 22 \pm 1 \  {\rm cm},
\end{equation}
neglecting small gravitational corrections.
Thus the probability of VCN reflection from the trap wall is equal to:
\begin{equation}
\label{probability}
P(v) = \frac{\tau^{VCN}_{st} \Delta x}{v} \left( 1+\epsilon \right), 
\end{equation}
where $\epsilon$ accounts for VCN losses in the entrance and exit windows and amounts to $\epsilon = 3.5\times10^{-4}$.
The probability of VCN reflection estimated in such a way for all 3 series of measurements is shown in Fig. \ref{reflectivity}.

This probability represents the reflectivity for the walls actually used in this experiment. 
It thus includes losses in the aluminium foil, and effects of the geometrical shape of the trap walls. 

On the other hand, we have in hand a model for the propagation of neutrons in the nanoparticle powder. 
The method used here to describe the reflection of VCN at a layer of nanoparticles is similar to that in \cite{Nesvizhevsky2008}. 
Namely, we neglected the relatively complex internal structure of the nanoparticle and modeled it as a uniform sphere. 
The neutron-nanoparticle elementary interaction was calculated using the first Born approximation. 
The amplitude for a neutron with the energy $(\hbar k)^2/(2m)$
to be scattered at a spherical nanoparticle with the radius R and Fermi potential V, at an angle $\theta$ is equal: 
\begin{equation}
\label{amplitude}
f(\theta) = - \frac{2m}{\hbar^2} \ V \ R^3 \ \left( \frac{\sin(q R)}{(q R)^3}-\frac{\cos(q R)}{(q R)^2} \right)
\end{equation}
where $q = 2 k \sin(\theta)$ is the transferred momentum.
The total elastic cross section is equal correspondingly:
\begin{equation}\label{xsel}
\sigma_{s} = \int \left| f \right|^2 d\Omega =  2 \pi
\left|\frac{2m}{\hbar^2} V \right|^2 \ R^6 \frac{1}{(k R)^2} I(k
R),
\end{equation}
where
\begin{equation}\label{Auxint}
I(k R) = \frac{1}{4}\left(1-\frac{1}{(2 k R)^2} +
                 \frac{\sin(4 k R)}{(2 k R)^3} - \frac{\sin^2(2 k R)}{(2 k R)^4}\right).
\end{equation}

The chemical composition of the nanoparticle is complex and includes carbon (up to 88\% of the total mass), 
hydrogen (1\%), nitrogen (2.5\%), oxygen (up to 10\%) \cite{Vereschagin}. 
Moreover, a certain amount of water covers a significant surface area of the nanoparticles. 
In general, the hydrogen in the water and on the surface of the nanoparticles scatters the neutrons up 
to the thermal energy range (\emph{up-scattering}); 
thermal neutrons do not interact as efficiently with nanoparticles and therefore traverse powder. 
The hydrogen quantity in the powder was measured by $(n,\gamma)$ method and compositions C$_{15}H$ and C$_{8}$H were found for degassed and non degassed powder respectively.

\begin{figure}
\begin{center}
\includegraphics[width=.9\linewidth]{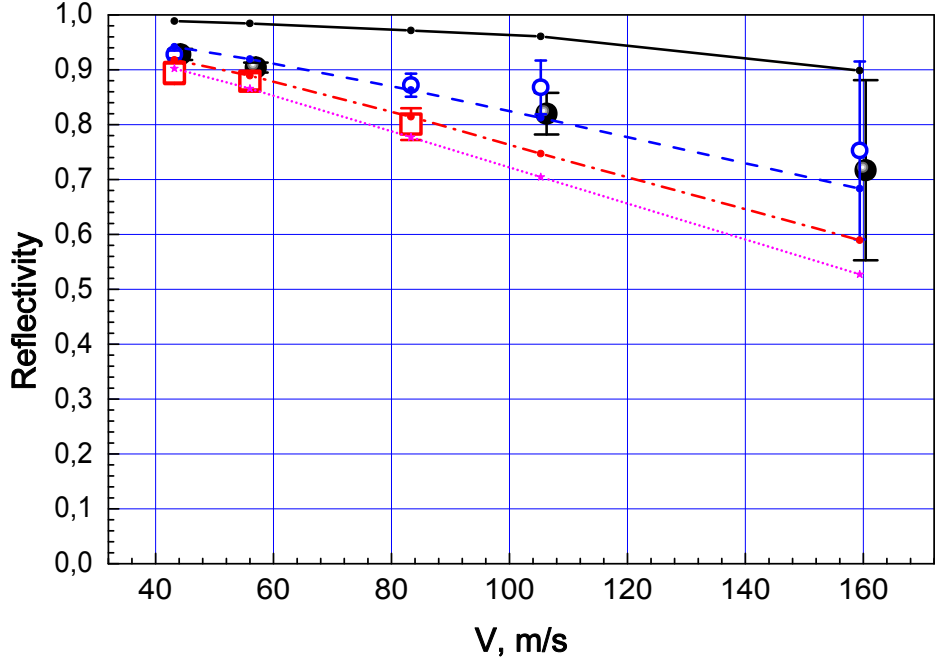}
\end{center}
\caption{
The probability of VCN reflection from a layer of nanoparticles as a function of their velocity.
Black circles correspond to measurements at room temperature after 12 hours pumping.
Red squares show measurements at room temperature after heating the trap at $120^\circ$~C in argon. 
Blue triangles indicate results obtained at a temperature of $150^\circ$~C under permanent pumping.
Thin lines correspond to Monte-Carlo calculations taking into account upscattering at hydrogen, 
the quoted cross section corresponds to neutron velocity of 100~m/s: 90 barn for dotted pink line, 60 barn for dot-dashed red line, 30 bar for dashed blue line and 0 barn for plain black line.
} \label{reflectivity2}
\end{figure}

To compare the experimental results to the model of independent nanoparticles, 
we further corrected the measured reflectivity for VCN loss in aluminum foils as well as for a dead zone in the wall structure between neighbor cylinders. 
The corrected reflectivity is shown in Fig. \ref{reflectivity2} for all 3 series of measurements; 
together with the Monte-Carlo calculation. 
The only parameter of this calculation concerns inelastic scattering on hydrogen. 
In the Monte-Carlo simulation, the quantity of hydrogen is fixed according to the C$_{17}$H chemical composition ($0.5$~\% of the total mass).
Besides, the cross section for inelastic scattering is assumed to follow the $1/v$ rule, 
and the value of the cross section $\sigma_H$ for neutron velocity of 100~m/s is considered as an effective parameter. 
The calculation reproduces well the velocity dependence of the reflectivity; the effective parameter is fitted at the approximate value of $\sigma_H = 30$~barn. 

\section{Conclusion}

We have observed for the first time storage of VCN with the velocity 40-160~m/s (the energy up to $10^{-4}$~eV) 
in a trap with walls containing powder of diamond nanoparticles. 
The VCN storage will allow us to accumulate significant number (density) of VCN in a trap (much larger than that typical for UCN). 
Such a trap can be used as a reflector for VCN and UCN sources. 
Further improvement of the VCN storage times could be achieved by removing a part of hydrogen from powder and by cooling a trap to a temperature, 
at which the inelastic up-scattering of VCN at hydrogen is suppressed. 
Another option could consist in replacing the diamond nanoparticles by O$_2$, D$_2$, D$_2$O, CO$_2$, CO 
or other low-absorbing nanoparticles, 
free of hydrogen and other impurities with significant VCN loss cross-section.


\end{document}